\documentclass[reprint,amsmath,amssymb,pra,superscriptaddress]{revtex4-2}
\usepackage{times}
\usepackage[pdftex]{graphicx}
\usepackage{dcolumn}
\usepackage{bm}
\usepackage{amsmath}
\usepackage{indentfirst}
\usepackage{float}
\usepackage[colorlinks,linkcolor=blue,anchorcolor=blue,citecolor=blue]{hyperref}
\usepackage[dvipsnames]{xcolor}
\usepackage{braket}% braket
\usepackage{multirow}
\usepackage{changes}
\usepackage{makecell}
\usepackage{amsthm,amssymb,bbm,newtxmath}
\usepackage{mathrsfs}
\usepackage[noend]{algpseudocode}
\usepackage{array}
\usepackage{booktabs}
\usepackage{multirow}
\usepackage{stackengine}

\usepackage{algorithmicx,algorithm}

\newcommand{\fid}{{F_{\text{avg}}}}

\newcommand{\x}[1]{{\text{#1}}}

\begin{document}

\title{Error-mitigated entanglement-assisted quantum process tomography}

\author{Zhihao~Wu}
\affiliation{College of Computer Science and Technology, National University of Defense Technology, Changsha 410073, China}

\author{Lingling~Lao}
\affiliation{College of Computer Science and Technology, National University of Defense Technology, Changsha 410073, China}

\author{Chengqi~Zhuke}
\affiliation{College of Computer Science and Technology, National University of Defense Technology, Changsha 410073, China}

\author{Yantong~Liu}
\affiliation{College of Computer Science and Technology, National University of Defense Technology, Changsha 410073, China}

\author{Xinfang~Zhang}
\affiliation{College of Computer Science and Technology, National University of Defense Technology, Changsha 410073, China}

\author{Shichuan~Xue}
\email{shichuanxue@quanta.org.cn}
\affiliation{College of Computer Science and Technology, National University of Defense Technology, Changsha 410073, China}

\author{Mingtang~Deng}
\affiliation{College of Computer Science and Technology, National University of Defense Technology, Changsha 410073, China}

\author{Junjie~Wu}
\affiliation{College of Computer Science and Technology, National University of Defense Technology, Changsha 410073, China}

\author{Kai~Lu}
\email{kailu@nudt.edu.cn}
\affiliation{College of Computer Science and Technology, National University of Defense Technology, Changsha 410073, China}
\begin{abstract}

In the era of noisy intermediate-scale quantum computing, it is of crucial importance to verify quantum processes and extract information. Quantum process tomography is a typical approach, however, both resource-intensive and vulnerable to state preparation and measurement errors. Here, we propose an error-mitigated entanglement-assisted quantum process tomography (EM-EAPT) framework to address these limitations. By leveraging a maximally entangled state to reduce state preparation complexity and integrating error mitigation techniques, our method significantly enhances robustness against SPAM errors. Experimental validation on a superconducting processor demonstrates the efficacy of EM-EAPT for two-qubit and three-qubit quantum processes. Results show more accurate average gate fidelities close to the realistic estimation, achieving $98.1\%\pm 0.03\%$ for a CNOT gate and $88.1\%\pm0.04\%$ for a cascaded CNOT process after error mitigation, compared to non-mitigated implementations. This work advances practical quantum verification tools for NISQ devices, enabling higher-fidelity characterization of quantum processes under realistic noise conditions.

\end{abstract}

\date{\today}
\maketitle

\address{}

\section{Introduction}

As quantum computing steps into the noisy intermediate-scale quantum (NISQ) era \cite{preskill2018quantum,brooks2019beyond,moll2018quantum}, we can develop high-precision and special-purpose quantum computers. Taking superconducting systems as examples, some representative quantum computing systems include Google's ``Sycamore" and ``Willow", USTC's ``Zuchongzhi" series, etc., scaling to over $100$ qubits \cite{acharya2024quantum, gao2024establishing,morvan2023phase,arute2019quantum,jurcevic2021demonstration,wu2021strong,cross2019validating}. 
Hence, it has become a core task to evaluate whether a quantum computing system has achieved the target quantum process in a noisy environment. For quantum computers, some verification methods are needed to determine whether quantum computing is correct and accurate. 

%Therefore, efficient and practical quantum verification and benchmarking methods have become increasingly important in the NISQ era. 

\begin{figure}
    \centering
    \includegraphics[width=0.8\linewidth]{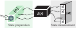}
    \caption{Quantum process tomography consisting of state preparation, quantum channel, and state measurement, where all components are influenced by unavoidable errors.}
    \label{fig:1}
\end{figure}

Quantum process tomography (QPT) is such a verification method \cite{ChuangNielsen1997,poyatos1997complete}. As shown in Fig.~\ref{fig:1}, QPT obtains the quantum channel information by inputting a series of probe states $\{\rho_i\}$ and measuring the observable values of these outputs $\{M_j\}$. It has the most complete and abundant quantum process information, providing the most comprehensive characterization of a quantum computing system, while, at the same time, they are inherently with the highest computational complexity. Besides, it is noted that the QPT method is inevitably influenced by the systematic errors and environmental noises, since it is a benchmarking tool that needed to be performed on a quantum system. 
Moreover, it requires exponential amounts of quantum state preparation and measurement, where both parts are unreliable and introduce tremendous errors. Considering the NISQ feature with environment noises and control errors, QPT is vulnerable to state preparation and measurement (SPAM) errors \cite{jackson2015detecting}, therefore, physically feasible only on the case up to three qubits from the perspective of experiment \cite{Bialczak2009Quantum,O2004Quantum,childs2001realization,shabani2011efficient,riebe2006process,govia2020bootstrapping,hou2020experimental}. Hence, it is of great significance to develop a robust and practical QPT method, facilitating the employment of QPT on NISQ computing systems.
%Various verification methods have been proposed for quantum computing systems, including quantum process tomography (QPT) \cite{ChuangNielsen1997,poyatos1997complete}, direct fidelity estimation \cite{flammia2011direct} randomized benchmarking \cite{knill2008randomized,magesan2011scalable}, cross-entropy verification \cite{boixo2018characterizing}, etc. Generally, QPT methods have the most complete and abundant quantum process information, providing the most comprehensive characterization of a quantum computing system, while, at the same time, they are inherently with the highest computational complexity. Randomized benchmarking obtains partial information about the system through some random gate sequences. In contrast, some verification methods, such as direct and indirect methods related to fidelity estimation and cross-entropy verification, can obtain less information while naturally possessing relatively low computational complexity. Besides, it is noted that QPT method requires exponential amounts of quantum state preparation and measurement, which introduces tremendous errors. So, it is vulnerable to state preparation and measurement (SPAM) errors \cite{jackson2015detecting}, therefore, physically feasible only on the case up to three qubits from the perspective of experiment \cite{Bialczak2009Quantum,O2004Quantum,childs2001realization,shabani2011efficient,riebe2006process,govia2020bootstrapping,hou2020experimental}.

On the other hand, entanglement-assisted process tomography (EAPT) \cite{altepeter2003ancilla,d2001quantum,leung2000towards,xue2022veapt} is an equivalent framework which utilizes the intrinsic relation between QPT and quantum state tomography (QST) based on Choi-Jamiolkowski isomorphism \cite{choi1975completely,leung2003choi} and imprints complete information about a quantum process on its output state \cite{d2003imprinting}. Compared to standard QPT, it consumes double amounts of qubits, generates one unique maximally entangled state involving two-qubit entangled operations, and introduces extra two-qubit control errors during state preparation. Thus, the experimental realization is realized on bulk optical platforms up to one-qubit cases \cite{altepeter2003ancilla,de2003exploiting}, showing its vulnerability towards state preparation errors.

% Error mitigation in quantum computing tasks is a crucial area of research aimed at enhancing the reliability and accuracy of quantum computations without the need for full error correction \cite{endo2018practical,endo2021hybrid,he2020zero}. These methods focus on improving the accuracy of quantum algorithms by addressing the noise and other imperfections inherent in quantum hardware, which are of great significance for NISQ computers that lack robust error correction capabilities. 
On the current NISQ quantum devices, error mitigation \cite{endo2018practical,lao2022software,cai2023quantum} is a crucial area of research aimed at enhancing the reliability and accuracy of quantum computations without the need for full error correction \cite{endo2018practical,endo2021hybrid,he2020zero}. Actually most applications, e.g., variational quantum algorithms \cite{cerezo2021variational}, involve applying short quantum circuits to some simple initial states for evolution and then estimating the expected values of certain observables. Although the quantum system undergoes coherent evolution, the effect of decoherence errors on estimating the expected values of observables is still evident. Therefore, we introduce the zero-noise extrapolation (ZNE) method, which can significantly improve the accuracy of the expected values in the presence of noise.

So in this paper, we put forward an error-mitigated entanglement-assisted process tomography (EM-EAPT). It borrows the idea from EAPT, which alleviates the exponential burden of state generation by fixing the input state as a maximally entangled state. Meanwhile, we combined the error mitigation approaches based on the simplified quantum circuits to enhance the robustness of SPAM errors. Experimentally, we demonstrated our method on two-qubit and three-qubit quantum processes on the superconducting quantum system. 

% In this paper, we combined the entanglement-assisted process tomography (EAPT) \cite{altepeter2003ancilla,d2001quantum,leung2000towards,xue2022veapt}  with typical error mitigation methods, which fixes the state preparation as a maximally entangled state, and facilitates the error mitigation method. To enhance the robustness of SPAM errors, we first introduced the EAPT method, which alleviated the exponential burden of state generation. Secondly, we combined the error mitigation approaches based on the simplified quantum circuits. Experimentally, we demonstrated our method on two-qubit and three-qubit quantum processes. Results validate the enhanced performance with the error mitigation technique.

This paper is organized as follows. In Sec.~\ref{sec:II}, we introduced the scheme of error-mitigated entanglement-assisted process tomography. Section~\ref{sec:IV} demonstrated our method on a superconducting chip focusing on two-qubit and three-qubit quantum processes. We concluded in Sec.~\ref{sec:V}.

\section{Error-mitigated entanglement-assisted process tomography}\label{sec:II}

% Specifically, standard QPT is a resource-demanding process due to the curse of dimensionality of both state preparation and measurement. There costs $4^n$ quantum states, $\{\rho_i\}=\{\ket{0},\ket{1},\ket{+},\ket{i}\}^{\otimes n}$, as probe states. Consequently, comprehensive quantum state tomography is necessary for each output state $\mathcal{E}(\rho_i)$, which requires $4^n$ independent measurements. In total, for a typical $n$-qubit quantum process, SQPT consumes $16^n$ measurements. 
% It can be seen that SQPT is resource-consuming from the perspective of state preparation, measurement, and post-processing. Meanwhile, it is vulnerable to SPAM error and difficult to scale to larger quantum systems since the number of state generation circuits is exponential.

\begin{figure}[hbtp]
    \centering
    \includegraphics[width=1\linewidth]{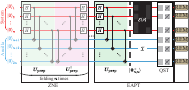}
    \caption{Error-mitigated entanglement-assisted quantum process tomography with zero-noise extrapolation. It contains the ZNE entanglement state preparation with different scaling factors, quantum channel, quantum state tomography and readout error mitigation.}
    \label{fig:2}
\end{figure}

EAPT is put forward in Ref.~\cite{choi1975completely,leung2003choi}. It imprints complete information about a quantum process on its output state via Choi-Jamiolkowski isomorphism. Compared to standard QPT, which costs $4^n$ quantum states, $\{\rho_i\}=\{\ket{0},\ket{1},\ket{+},\ket{i}\}^{\otimes n}$, as probe states, and corresponding quantum state tomography for each output state, EAPT is an equivalent alternative in terms of the total measurements.
Specifically for an $n$-qubit quantum process, EAPT needs to introduce $n$ ancillary qubits and construct one $2n$-qubit maximally entangled state 
\begin{equation}
    \ket{\Phi_{n,n}^+}=(1/\sqrt{d})\sum_{j=1}^d \ket{j}\otimes \ket{j}.
\end{equation}
Then, we subject the system space to the channel of $\mathcal{E}(\rho)$, keeping the ancillary space identity. We can obtain the joint output density matrix $\rho_{\x{out}}=\mathcal{E}\otimes I (\ket{\Phi_{n,n}^+}\bra{\Phi_{n,n}^+})$ by performing a full QST. The overall EAPT scheme is illustrated in the line box of Fig.~\ref{fig:2}. After an eigen-decomposition, we have $\rho_{\x{out}}=\sum_k\ket{a_k}\bra{a_k}$. Divide $\ket{a_k}$ into $n$ equal segments, then, $A_k$ is the $n\times n$ matrix having the $i$-th segment as its $i$-th column. So we get the quantum process Kraus-operator sum form $\mathcal{E}(\rho)=\Sigma_kA_kMA_k^\dagger$. 

The EAPT method shifts the complexity from state preparation to measurement, which consumes the same amount of measurement in total, but only one unique state preparation circuit. However, it introduces more vulnerable two-qubit gates and generates the entangled state $\ket{\Phi_{n,n}^+}$, thus, naturally more sensitive to state preparation errors.
On the other hand, since the state preparation process is extremely simplified to only one maximally entangled state, it is possible to combine some error mitigation methods to enhance the performance of the EAPT method.
% \begin{table}[hbtp]
% \label{tab.2}
% \centering
%     \caption{Comparison between SQPT and EAPT}
%     \label{Tab:1}
% \tabcolsep=0.3cm
% \renewcommand{\arraystretch}{1.5}
% \begin{tabular}{c|c|c|c}
% \hline
%      & \# of qubits & \# of inputs & \# of meas.    \\ \hline
% SQPT & $n $           & $4^n$        & $4^n\times 4^n=16^n$  \\ 
% EAPT & $2n$           & $\mathbf{1}$          & $16^n$ \\ \hline
% \end{tabular}
% \end{table}

Therefore, this paper attempts to realize an error-mitigated EAPT method on superconducting qubits, showing its robustness towards complicated realistic noise conditions.

It can be seen that in the EAPT scheme, the core component is the state generation circuit $U_\x{prep}$ and the state tomography. Generally for QST, we utilize the maximally likelihood estimation to guarantee a physical quantum state density matrix (details in the Appendix~\ref{app:qst}). The goal of the optimization is to maximize the following likelihood function
\begin{equation}
\label{eq.mle}
   \mathcal{L}(t) = \frac{1}{c}\prod_i\x{exp}\left[-\frac{(\langle\psi_{i}|\hat{\rho}(t)|\psi_{i}\rangle - p_{i})^2}{2\langle\psi_{i}|\hat{\rho}(t)|\psi_{i}\rangle}\right],
\end{equation}
where $c$ is the normalization factor, $\hat{\rho}(t)$ is the optimizing target, $\langle\psi_{i}|\hat{\rho}(t)|\psi_{i}\rangle$ are the ideal measurement results and $p_{i}$ are the experimental values. Considering the SPAM errors in the system,  $p_{i}$ are easily influenced. Thus, we employ the typical error mitigation method to reduce the errors during entanglement state preparation and readout (REM), as shown in the dotted box in Fig.~\ref{fig:2}.

%% extrapolation circuits construction
The main idea of the zero-noise extrapolation method is to calculate the expected values of observables under different noise intensities and then use the linear extrapolation method to offset the influence of different orders of noise on the calculation of expected values, thereby extrapolating to the expected value when the noise is limited to zero.

To apply the zero-noise extrapolation method, we need to artificially scale the noise of the quantum circuit to obtain the expected values of observables under different noise levels. In most cases, we cannot directly control the noise of the physical hardware. Still, we can indirectly increase the noise by manipulating the circuit while keeping the circuit logic unchanged. Here, we utilize the typical unitary folding method to achieve noise scaling.

Unitary folding is implemented by mapping $U_{\x{prep}} \to \left(U_{\x{prep}}U_{\x{prep}}^{\dagger}\right)^nU_{\x{prep}}$ on the quantum gate and $n$ is the circuit folding times. Such mappings can be performed globally or locally. As shown in Figure~\ref{fig:2}, we utilize the global mapping that acts on the entire circuit. The new circuit thus generated has the same evolution effect on the initial state as the original circuit. Still, the number of quantum gates is three times that of the original. Therefore, the noise is equivalent to being amplified three times the original.

% For SQPT, it is almost impractical to perform ZNE to eliminate state preparation errors since there exist exponential input states. Regarding EAPT, however, it is naturally suitable to conduct such error mitigation trials.

%%    QST measurement value extrapolation
    
To sum up, we combine the ZNE method with EAPT since only one state preparation is needed. First, we fold the state preparation circuit more times to obtain the scaling circuits, as shown in Fig.~\ref{fig:2}. Then, for each scaling circuit, we conduct a full quantum state tomography on the output state, obtaining the measurement values on different basis.
Third, we employ the linear extrapolation method on the measurement data of different scaling circuits to get the desired error-mitigated zero-noise measurement data $p_{i}^{\x{EM}}$. Thus, the maximally likelihood estimation target changes to

\begin{align}
      \mathcal{L}(t)^{\x{EM}} &= \frac{1}{c}\prod_i\x{exp}\left[-\frac{(\langle\psi_{i}|\hat{\rho}(t)|\psi_{i}\rangle - p_{i}^{\x{EM}})^2}{2\langle\psi_{i}|\hat{\rho}(t)|\psi_{i}\rangle}\right],\\
   p_{i}^{\x{EM}} &= \x{ZNE}\left(p_{i}^{s_1},p_{i}^{s_2},\ldots,p_{i}^{s_i},\ldots\right).
\end{align}

\section{Experimental realization of EM-EAPT on superconducting qubits}\label{sec:IV}

% \begin{figure*}[hbtp]
%     \includegraphics[width=1\linewidth]{figure3.pdf}
%     \caption{Experimental realization of EAPT with error mitigation on superconducting qubits. (a) Device sketch with readout resonators, transmon qubits, and electronic controls. (b) Selected qubits with three system qubits (red crosses) and three ancillaries (blue crosses). The zoom-in picture is the tunable coupling architecture. (C) Two-qubit entanglement state fidelity $\mathcal{F}$ among $\x{S}_1-\x{S}_2$ (red diamonds) and $\x{S}_2-\x{S}_3$ (blue rectangles). (d) Average $\x{CNOT}$ quantum process fidelity $\fid$ with different scaling circuits and linear extrapolation circuits. (e) Six-qubit entanglement state density matrices (real matrix at top and imaginary matrix at bottom. The three density matrices are the ideal one $\ket{\Phi_{3,3}^+}\bra{\Phi_{3,3}^+}$, the linear extrapolation one $\rho_{\x{lin}}$ ($\mathcal{F}=96.94\%$) and the original $1\x{X}$ circuit one $\rho_{\x{exp}}$ ($\mathcal{F}=89.30\%$). (f) Average $\x{CCNOT}$ quantum process fidelity $\fid\left( \x{CCNOT}_{\x{S}_1\x{S}_2\x{S}_3}\right)$ with different scaling circuits and linear extrapolation circuits.}
%     \label{fig:3}
% \end{figure*}

To demonstrate the EM-EAPT method experimentally, we select $6$ coupled qubits from a %$66$-qubit 
2D superconducting quantum processor (with flux-tunable transmon qubit and flux-tunable transmon coupler~\cite{arute2019quantum, wu2021strong, zhu2022quantum}).
Among the selected qubits, three qubits ($S_1$, $S_2$, and $S_3$) are used as system qubits to perform target processes to be analyzed, and the other three qubits ($A_1$, $A_2$, and $A_3$) are ancilla qubits to be entangled with the system.
The Rxy gate, CZ gate, and readout fidelities of these elements are shown in Tab.~\ref{tab:2}.
More details about device setting and parameter calibration are described in Appendix~\ref{app:characterization}.

\begin{table}[htbp]
\label{tab:2}
\centering
\caption{Featured operation error rates of the selected elements.}
\begin{tabular}{lcccccc}
\toprule
\multicolumn{1}{l}{Qubit} & S$_{1}$ & S$_{2}$ & S$_{3}$ & A$_{1}$ & A$_{2}$ & A$_{3}$  \\
\midrule
Readout error (\%) & 3.7 & 5.4 & 3.5 & 4.8 & 2.6 & 4.3  \\
Rxy gate error (\%) & 0.11 & 0.10 & 0.11 & 0.16 & 0.16 & 0.11  \\
\midrule
CZ gate & $\x{C}_{\x{S}_1\x{A}_1}$ & $\x{C}_{\x{S}_2\x{A}_2}$ & $\x{C}_{\x{S}_3\x{A}_3}$ & $\x{C}_{\x{S}_1\x{S}_2}$  & $\x{C}_{\x{S}_2\x{S}_3}$  & - \\
\midrule
CZ gate error (\%) & 1.19 & 3.32 & 0.80 & 2.17 & 2.21 & -  \\
\bottomrule
\end{tabular}
\end{table}

\begin{figure}[hbtp]
    \centering
    \includegraphics[width=1\linewidth]{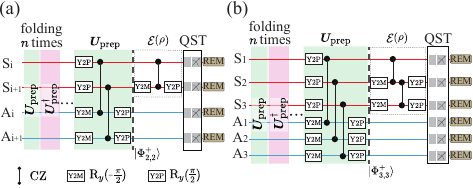}
    \caption{Experimental device and circuits. 
    %(a) Selected qubits with three system qubits (red crosses) and three ancillaries (blue crosses). The zoom-in picture is the tunable coupling architecture. 
    (a) Two-qubit EM-EAPT circuits with CNOT quantum process as target. (b) Three-qubit EM-EAPT circuits with cascaded CNOTs quantum process as target. }
    \label{fig:3}
\end{figure}

To show the contribution of our method to the development of quantum computing, here we apply it to research on the noise of two meaningful target processes:
\begin{enumerate}
    \item A single CNOT gate, which is a necessary basic building elements for various quantum algorithms.
    \item Cascaded two CNOT gates with a common control qubit, which is critical for the error syndrome of surface code~\cite{fowler2012SurfaceCodesPractical}.
\end{enumerate}
Fig.~\ref{fig:3}(a) and (b) plot the circuits of the above two-qubit and three-qubit EM-EAPT, respectively. The state preparation circuit $U_\x{prep}$ generates the maximally entangled state $\ket{\Phi_{n,n}^+}$, containing $R_y$ rotational gates on each qubit and $\x{CNOT}$ gates with $\x{Y2M} \cdot \x{CZ} \cdot \x{Y2P}$, where 
\begin{align}
\nonumber R_{y}(\theta)=\left[\begin{array}{rr}
\cos \frac{\theta}{2} & -\sin \frac{\theta}{2} \vspace{0.5ex}\\
\sin \frac{\theta}{2} & \cos \frac{\theta}{2}
\end{array}\right],\\ \x{Y2P}=R_y(\frac{\pi}{2}),\x{ } \x{Y2M}=R_y(-\frac{\pi}{2}).
\end{align}

%Figure~\ref{fig:3}(a) shows the core devices, including readout units, transmon qubits, and electronic controls.
%We utilize microwave signals to control qubit XY rotations, flux bias lines to adjust qubit frequency and tune the couplers, and dispersive measurement scheme with Josephson parameter amplifiers to readout qubit status.
%On the chip, the superconducting qubits are frequency-tunable transmon qubits (cross labels in the schematic), interconnected by a tunable coupler qubit between adjacent transmon qubits (grey rectangular).
% Specifically, we choose six qubits in Fig.~\ref{fig:3}(b) as the research targets (three system qubits, three ancilla qubits, and corresponding couplers). The three system qubits $\text{S}_1, \text{S}_2, \text{S}_3$ are connected one by one so that we can study a fully entangled process. For each system qubit, we select a corresponding ancilla $\text{A}_1, \text{A}_2, \text{A}_3$ to generate the entanglement pair. For these six qubits, we summarized the relative typical parameters in Tab.~\ref{tab.4}.

As stated in Sec.~\ref{sec:II}, we construct the EAPT circuit, conduct QST on the output, and rebuild the quantum process. Also, we employ the ZNE method during entanglement state preparation. 

To verify the ZNE results, we evaluate the entanglement state fidelity $\mathcal{F}$ and the final quantum process average fidelity $F_{\text{avg}}$. For $\mathcal{F}$, we conduct the different scaling circuits, perform the QST, rebuild the output state density matrix, and calculate the state fidelity based on the mitigated state
\begin{equation}
\mathcal{F}(\rho, \sigma) = \left( \text{Tr} \sqrt{\sqrt{\rho} \sigma \sqrt{\rho}} \right)^2.
\end{equation}
We could demonstrate the efficacy of the ZNE error mitigation method on state preparation. For $F_{\text{avg}}$, we add the $\mathcal{E}\otimes I$ process after the entanglement state generation circuit, employ the ZNE method to get mitigated output choi state, and rebuild the quantum process. The average quantum process fidelity  $F_{\text{avg}}$ between the rebuilt $\mathcal{U}$ and the real $\mathcal{E}$ is given by 
\begin{equation}
\label{eq.favg}
    \fid(\mathcal{U}, \mathcal{E}) =\int d \psi\left\langle\psi\left|\mathcal{U}^{\dagger} \mathcal{E}(|\psi\rangle\langle\psi|) \mathcal{U}\right| \psi\right\rangle, 
\end{equation}
where the integral is taken according to the uniform Haar-invariant probability measure on state vectors $\ket{\psi}$. In fact, the average fidelity of the quantum gate $\fid(\mathcal{U}, \mathcal{E})$ has a relationship with the diamond norm $\|\mathcal{U}- \mathcal{E}\|_\diamond$, which is generally used to compare different quantum channels. For simplicity, we utilize the average quantum gate fidelity $\fid$ to characterize the numerical simulation results.

\begin{figure}[htbp]
    \centering
    \includegraphics[width=1\linewidth]{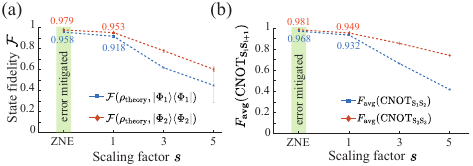}
    \caption{Two-qubit EM-EAPT results. (a) Two-qubit entanglement state fidelity $\mathcal{F}$ among $\x{S}_1$-$\x{S}_2$ (red diamonds) and $\x{S}_2$-$\x{S}_3$ (blue rectangles). (b) Average $\x{CNOT}$ quantum process fidelity $\fid$ with different scaling factor $s$ circuits and the linear extrapolation circuits (green shaded).}
    \label{fig:4}
\end{figure}

Taking two-qubit CNOT gate as the target quantum process, we first check the results of ZNE on state preparation. 
Figure~\ref{fig:4}(a) shows the entanglement state generation fidelities with different scaling circuits (the horizontal axis scaling factor $s$ corresponds to the folding times $n$ with $s=2n+1$), where $\mathcal{F} = \mathcal{F}(\rho_\x{theory}, \ket{\Phi_i}\bra{\Phi_i})$, and $\rho_\x{theory}=\ket{\Phi_{n,n}^+}\bra{\Phi_{n,n}^+}$ and $\ket{\Phi_i}$ is the $i$-th generated state as shown in Fig.~\ref{fig:3}(b). Utilizing the linear extrapolation method, we can see that the entanglement state fidelity has been enhanced from the original $95.3\%\pm 0.5\%$ and $91.8\%\pm 1.1\%$ to $97.9\%\pm0.2\%$ and $95.8\%\pm 0.2\%$. 

%% Improvement of the process fidelity
Based on the mitigated entangled state, we consecutively operate the target quantum process $\x{CNOT}\otimes I$ and perform the full QST on the output. Via the maximal likelihood estimation method (details in Appendix~\ref{app:qst}), we could obtain the most likely physical density matrix $\rho_{\text{out}}$. And through eigen-decomposition, we arrive at the quantum process's Kraus operator sum form. We evaluate the average quantum process fidelity following Eq.~(\ref{eq.favg}), and Fig.~\ref{fig:4}(b) shows the $F_{\x{avg}}$ with the scaling factor of circuits. Average quantum gate fidelities are raised to $98.1\% \pm 0.03\%$ and $96.8\% \pm 0.04\%$, respectively, compared to the naive EAPT method.

\begin{figure}[htbp]
    \centering
    \includegraphics[width=1\linewidth]{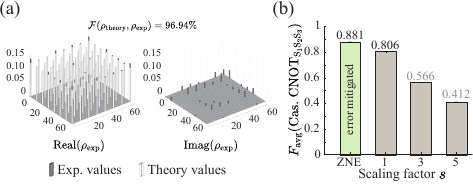}
    \caption{Three-qubit EM-EAPT results. (a) 
    A typical six-qubit error-mitigated entanglement state $\rho_{\x{exp}}$ density matrices compared with the ideal state $\rho_{\x{theory}}=\ket{\Phi_{3,3}^+}\bra{\Phi_{3,3}^+}$ (real matrix at left and imaginary matrix at right, gray boxes for experimental values and dotted boxes for theory values). (b) Average quantum process fidelity $\fid\left( \x{Cas. CNOT}_{\x{S}_1\x{S}_2\x{S}_3}\right)$ with different scaling factor $s$ circuits and the linear extrapolation circuits (green shaded).}
    \label{fig:5}
\end{figure}

Further, we validate our method on a six-qubit experiment aiming at a three-qubit cascaded CNOTs process. Figure~\ref{fig:5}(a) illustrates the generated error-mitigated six-qubit entanglement state matrix with the state fidelity $\mathcal{F}=96.94\%$, compared to the original circuit with $\mathcal{F}=89.3\%$. And Fig.~\ref{fig:5}(b) shows the enhanced average quantum process fidelity from original EAPT $\fid=80.6\%\pm 0.03\%$ to linear extrapolation $\fid=88.1\%\pm 0.04\%$.

Results show that the EM-EAPT method augments the performance of the QPT task, obviously, validating the ZNE effects of error mitigation.

\section{Conclusion}\label{sec:V}
In this work, we proposed an error-mitigated entanglement-assisted quantum process tomography (EM-EAPT) method and demonstrated on superconducting qubits, addressing critical challenges in resource efficiency and noise resilience.  
By simplifying the state preparation to a single maximally entangled state, it becomes possible to implement error mitigation strategies that are difficult to apply in traditional QPT with exponential state generation circuits.
Experimental results on two-qubit CNOT and three-qubit cascaded CNOT processes demonstrate fidelity improvements of up to $98.1\%$ and $90\%$, respectively, highlighting the effectiveness of the approach. The framework not only provides a practical tool for characterizing quantum processes in NISQ systems but also opens avenues for integrating advanced error mitigation strategies into quantum benchmarking protocols. 
%In this work, we combined EAPT with error mitigation methods for quantum process tomography on superconducting qubits. By simplifying the state preparation process in EAPT to a single maximally entangled state, it becomes possible to implement error mitigation strategies that are difficult to apply in SQPT. The experimental validation on two-qubit and three-qubit systems shows that these error mitigation methods can effectively improve the performance of quantum process tomography. Specifically, the ZNE method significantly enhances the entanglement state fidelity and the average quantum process fidelity. The results of this work provide a practical and robust way to benchmark and verify quantum processes, which is of great significance for the development of quantum computing in the NISQ era.

\vspace{5pt}
\textbf{Acknowledgments.} We appreciate the helpful discussion with other members of the \texttt{QUANTA} group, including Xiang Fu, Anqi Huang, Shun Hu, Cong Li, Yang Yang, Xiaofeng Yi, Yichuan Zeng, and Weichen Wang. We also thank Shaojun Guo and Wenhao Chu for the fruitful discussion. This work was supported by the National Natural Science Foundation of China under Grant No. 62301572. 

\vspace{5pt}
\textbf{Contribution.} Z.-H.W. ,X.-F.Z., and S.-C.X. conducted the qubit experiments; Z.-H.W., S.-C.X., C.-Q.Z.-K., and L.-L.L. analyzed the experimental data; L.-L.L. and Y.-T.L. developed the error mitigation method; M.-T.D., J.-J.W, and  K.L. supervised both the theory and experiment. All authors contributed to the result discussion and manuscript writing.

\appendix

\section{Standard quantum state tomography with maximally likelihood estimation}\label{app:qst}

Theoretically, we can decompose the rebuilt output state as the combination of probe states, 
    \begin{equation}
    \label{eq:1}
    \mathcal{E}\left(\rho_{j}\right)=\sum_{k} \lambda_{jk} \rho_{k},
    \end{equation}
    where the matrix $\lambda_{jk}$ can be determined. Moreover, the quantum process can be expressed as the operator-sum formula 
    \begin{equation}
        \mathcal{E}(\rho) = \sum_{m,n = 0}^{d^2 - 1} \chi_{mn} \hat{A}_m \rho \hat{A}_n^{\dagger}.
    \end{equation}
    
Together with Eq. (\ref{eq:1}), we can give the $\chi$ matrix by inverting the $\beta$ matrix. Solving the inverse problem of state estimation involves in huge computational costs \cite{banaszek1999maximum,ferrie2014self}. However, standard reconstruction techniques typically lead to an unphysical process matrix because of noise and error. We can use maximum likelihood estimation to find a physical process matrix as close as possible to the experimental data.

Any $n$-qubit density matrix $\hat{\rho}$ can be uniquely represented as
\begin{equation}
\label{eq.a1}
    \hat{\rho}=\frac{1}{2^{n}}\sum_{i_{1},i_{2},\cdots,i_{n}=0}^{3}S_{i_{1},i_{2},\cdots,i_{n}}\hat{\sigma}_{i_{1}}\otimes\hat{\sigma}_{i_{2}}\otimes\cdots\otimes\hat{\sigma}_{i_{n}}.
\end{equation}
The $\hat{\sigma}_{i}$ matrices are
\begin{equation}
\hat{\sigma}_0=I,\hat{\sigma}_1=\begin{pmatrix}
0 & 1\\
1 & 0
\end{pmatrix},\hat{\sigma}_2=\begin{pmatrix}
0 & -i\\
i & 0
\end{pmatrix},\hat{\sigma}_3=\begin{pmatrix}
1 & 0\\
0 & -1
\end{pmatrix}.
\end{equation}

The coefficients $S_{i_{1},i_{2},\cdots,i_{n}}$ are the Stokes parameters, and normalization requires that $S_{0,0,\ldots 0} = 1$, allowing $4^n - 1$ real parameters. Theoretically, the $S_{i}$ can be determined by experimental measurement $S_i=\mathrm{Tr}\left\{\hat{\sigma}_i\hat{\rho}\right\}$ ,thus, deriving the state matrix $\hat{\rho}$ as in Eq.~\ref{eq.a1}. However, since the Stokes parameters are calculated based on experimental values, the
statistical fluctuations or drifts often yield nonphysical results, which means that
the density matrix calculated might not be Hermitian, positive, semi-definite, and normalized. To tackle this problem, we typically utilize the maximum likelihood tomographic technique for estimating the quantum state, optimizing a physical density matrix most likely to the measured data. 

To guarantee a physical quantum state matrix, we consider the density matrix that can be written as $\mathbf{T}^\dagger\mathbf{T}$, which is both positive semi-definite and Hermitian. Along with the normalization by its trace $\mathbf{T}^\dagger\mathbf{T}/\mathbf{T}\mathbf{r}(\mathbf{T}^\dagger\mathbf{T})$, we obtain a legitimate physical density matrix $\rho(t)=\mathbf{T}^\dagger(\mathbf{t})\mathbf{T}(\mathbf{t})/\mathrm{Tr}(\mathbf{T}^\dagger(\mathbf{t})\mathbf{T}(\mathbf{t}))$, where $\mathbf{T}(\mathbf{t})$ follows the invertible form:
\begin{equation}\mathbf{T}(\mathbf{t}):=
\begin{pmatrix}
t_1 & 0 & 0 & 0 & 0 \\
t_{n+1}+it_{n+2} & t_2 & 0 & 0 & 0\\
t_{3n-1}+it_{3n} & t_{n+3}+it_{n+4} & t_3 & 0 & 0\\
\vdots & \vdots & \cdots & \ddots & 0 \\
t_{n^2-1}+it_{n^2} & \cdots & \cdots& \cdots & t_n
\end{pmatrix}
\end{equation}
Thus, the optimization problem is reduced to finding the minimum of the function in Eq.(\ref{eq.mle}).  In practice, we employ the Python built-in convex optimization package cvxpy to perform the minimization process.

\section{Experimental details}
\label{app:characterization}

\subsection{Device setting}

% processor information
\begin{figure*}[htbp]
    \centering
    \includegraphics[width=0.81\linewidth]{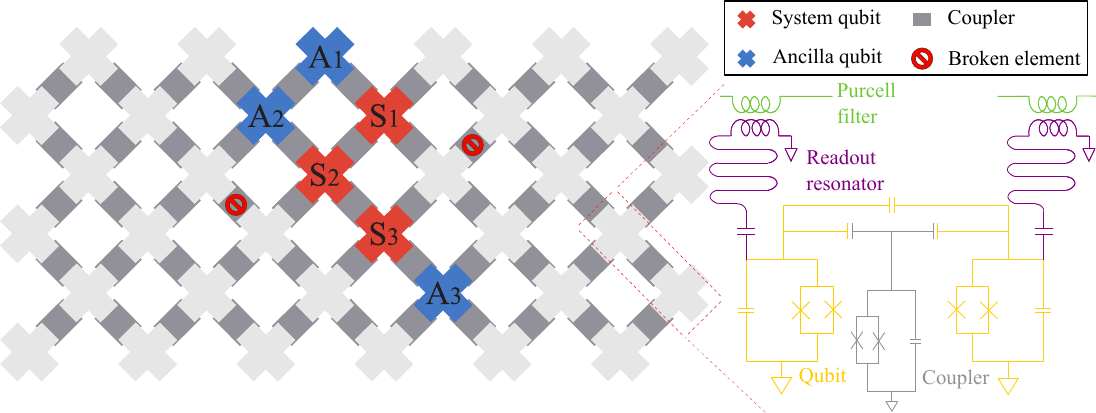}
    \caption{The schematic diagram of the quantum processor. The flux-tunable transmon qubits form a 2D mesh while each pair of the neighboring qubits are coupled by a flux-tunable transmon coupler. Each of the qubit is coupled with a readout resonator for dispersive readout. The resonators on the same row shares a common Purcell filter for multiplexed readout. The selected region for our experimental demonstration are highlighted. The broken couplers, which are flux untunable or induce chip heating, are also marked in the graph.}
    \label{fig:processor-information}
\end{figure*}

Our experiment is performed on a %$66$-qubit 
2D superconducting quantum processor, which employs an architecture of flux-tunable transmon qubits, flux-tunable transmon couplers, and resonator-based dispersive readout~\cite{arute2019quantum, wu2021strong, zhu2022quantum}.
The schematic diagram of our processor is shown in Fig. ~\ref{fig:processor-information}.

Among all the elements on the processor, we chose a region covering 6 qubits for our experimental demonstration, considering various factors (featured parameter values, operation fidelity, qubit decoherence, crosstalk strength, chip topology, etc.), as shown in Fig. ~\ref{fig:processor-information}.

% cryogenic environment

The cryogenic environment of our processor is provided by a dilution refrigerator.
% , as shown in Fig. ~\ref{fig:cryo-env}.
The temperature of the MC plate was kept around 20 MK during our experiment.
The connection between the processor and our room-temperature electronics is mediated by attenuators, filters, and amplifiers.
The attenuators and filters are designed to suppress signal noise while leaving enough signal power.
The Josephson parametric amplifier (JPA) at each readout signal output line is used to improve readout SNR, and subsequent high-electron mobility transistor (HEMT) and room temperature microwave amplifier to enhance signal power.

% \begin{figure}[hbtp]
%     \includegraphics[width=0.91\linewidth]{CryoSetup.pdf}
%     \caption{The cryogenic environment of the quantum processor.}
%     \label{fig:cryo-env}
% \end{figure}

% control electronics & software

We use integrated room-temperature signal generation and acquisition electronics to calibrate this chip.
The DC fluxbias and the flux pulse for each transmon are provided by a single Z channel.
The XY microwave signal for each qubit is generated by an XY channel, which employs a numerical controlled oscillator (NCO) as LO and a digital mixer for frequency up-conversion, removing the LO leakage and mirror sideband noises.
The multiplexed readout of 6 qubits on a single readout line is realized by a single pair of RO-IN and RO-OUT channels.

We also utilize high-level management software to schedule hardware resources, perform standardized calibrations, manage parameters, and execute target algorithms through well-defined interfaces.

\subsection{Parameters calibration}

% 比他们大文章写法，更简单，不写都做了什么及其逻辑，只写哪些参数关键

Here we list the most significant parameters for our experiment.
Except for the early-stage elements aliveness check, several highly dimensional parameter optimization, and final-stage fine tuning, calibrations, and maintenance for the parameters are organized by our automatic calibration framework QoPilot~\cite{wu2025qopilot}.
More details about our calibration design, their scheduling, and parameter refreshment are also described in Ref. ~\cite{wu2025qopilot}.

\begin{table*}[htbp]
\label{tab.app}
\centering
\caption{Featured parameters of the selected elements.}
\begin{tabular}{lcccccc}
\toprule
\multicolumn{1}{l}{Qubit} & S$_{1}$ & S$_{2}$ & S$_{3}$ & A$_{1}$ & A$_{2}$ & A$_{3}$  \\
\midrule
Idle freq. (GHz) & 4.490 & 4.611 & 4.577 & 4.565 & 4.414 & 4.393  \\
Anharmonicity (MHz) & -245 & -247 & -250 & -257 & -248 & -248 \\
F01$_\text{max}$ (GHz) & 4.648 & 4.633 & 4.678 & 4.689 & 4.556 & 4.477 \\
Qubit freq during readout (GHz) & 4.4 & idle freq. & idle freq. & idle freq. & idle freq. & idle freq. \\
Readout freq. (GHz) & 6.431 & 6.406 & 6.435 & 6.403 & 6.372 & 6.346 \\
T1 idle ($\mu$s) & 39.5 & 56.6 & 36.0 & 45.5 & 28.4 & 47.2 \\
T2$^*$ idle ($\mu$s) & 4.0 & 7.1 & 3.8 & 4.7 & 3.7 & 5.3 \\
T1 avg. ($\mu$s) & 34.5 & 33.4 & 34.3 & 36.9 & 33.3 & 34.7 \\
T2$^*$ avg. ($\mu$s) & 4.5 & 4.3 & 3.3 & 4.6 & 4.1 & 4 \\
\midrule
Coupler & $\x{C}_{\x{S}_1\x{A}_1}$ & $\x{C}_{\x{S}_2\x{A}_2}$ & $\x{C}_{\x{S}_3\x{A}_3}$ & $\x{C}_{\x{S}_1\x{S}_2}$ & $\x{C}_{\x{S}_2\x{S}_3}$ & - \\
\midrule
Interaction freq. (GHz) & (4.405, 4.660) & (4.605, 4.363) & (4.570, 4.325) & (4.375, 4.600) & (4.510, 4.262) & -  \\
\bottomrule
\end{tabular}
\end{table*}

\subsubsection{Qubit parameters}

\begin{itemize}
\item Qubit idle frequency. Its value is set by our frequency allocator, as shown in Tab.~\ref{tab.app}.
\item Qubit anharmonicity.
\item Qubit F01 spectra.
\item $\pi/2$ pulse parameters, including pulse length, pulse amplitude, and DRAG coefficient~\cite{motzoi2009SimplePulsesElimination, gambetta2011AnalyticControlMethods}. These parameters describe $\pi/2$ pulse with cosine envelope, and any $\pi$ pulse is realized by cascaded two $\pi/2$ pulses. Pulse length is fixed at 40 ns to eliminate leakage error, and thus our DRAG correction focuses on handling phase error.
\item Qubit Z pulse distortion compensation coefficients~\cite{yan2019StronglyCorrelatedQuantum}. For simplicity, in this experiment we focus on long-range exponential tailing.  The Z pulse nonlinear response is also under consideration.
\item Qubit decoherence parameters, including T1 idle, T2 idle, T1 spectra, and T2 spectra.
\item XEB1Q fidelity~\cite{boixo2018characterizing, arute2019quantum, wu2021strong, zhu2022quantum}.
\end{itemize}

\subsubsection{Readout parameters}

\begin{itemize}
\item Readout pulse parameters, including readout frequency, readout amplitude, and readout length. The readout power is finely suppressed to avoid AC Stark shift induced simultaneous readout error. Here we simply use square integration filter, whose length is identical with readout pulse.
\item Qubit frequency during readout. Its value could be deviated from qubit idle frequency to avoid frequency collision due to AC Stark shift during readout~\cite{arute2019quantum, wu2021strong}, as shown in Tab.~\ref{tab.app}.
\item Readout IQ cloud parameters, including 0/1 IQ discrimination level and readout fidelities, based on which qubit-individual readout error correction~\cite{yan2019StronglyCorrelatedQuantum} has been applied during parameters calibration and algorithm execution.
\item JPA control parameters, including pump amplitude, pump frequency, and fluxbias amplitude.
\end{itemize}

\subsubsection{Coupler and CZ parameters}

\begin{itemize}
\item Coupler turning-off point. The surrounding qubits on the processor are decoupled from the selected qubits by turning off the in-between couplers.
\item Coupler's coupling strength function $g(\Phi_c)$.
\item Adiabatic CZ~\cite{google2020DemonstratingContinuousSet, sung2021RealizationHighFidelityCZ} (ACZ) pulse parameters, including qubit detuning frequencies, coupling strength turned on, pulse length, padding length, etc. Here we fix pulse length at 40 ns. Interaction frequencies are determined by our frequency allocator, as shown in Tab.~\ref{tab.app}. The padding length after ACZ pulses is chosen to avoid the influence of residual short-range qubit Z pulse tailing.
\item Dynamic phases of ACZ.
\item CZXEB fidelity~\cite{boixo2018characterizing, arute2019quantum, wu2021strong, zhu2022quantum}.
\end{itemize}

\subsubsection{Signal delay parameters}

\begin{itemize}
\item QubitXY-QubitZ delay.
\item QubitXY-CouplerZ delay.
\item Readout acquisition delay.
\end{itemize}

%REF
\bibliographystyle{apsrev4-1}
\bibliography{refs}

\end{document}